# Security in Drones

Jonathan Morgan[1] , Julio Perez[2], Jordan Wade[3], Sundar Krishnan[4]

Dept. of Computer Science, Angelo State University

[1]jmorgan28@angelo.edu, [2]jperez146@angelo.edu , [3]jwade9@angelo.edu , [4]skrishnan@angelo.edu

*Abstract*—**Drones are used in our everyday world for private, commercial, and government uses. It is important to establish both the cyber threats drone users face and security practices to combat those threats. Privacy will always be the main concern when using drones. Protecting information legally collected on drones and protecting people from the illegal collection of their data are topics that security professionals should consider before their organization uses drones. In this article, the authors discuss the importance of security in drones.**

*Keywords—drones, technology, security, privacy, information*

## I. INTRODUCTION

Today in society we surround ourselves with technology ranging from automobiles to cell phones. A technology many people are not familiar with is the drone. Today many companies are using drones for various activities such as performing search and rescue, delivering groceries to your doorstep, and taking amazing photos. With anything, there are also cons to using drones, the main one being that they are susceptible to being hacked. Drones being hacked can cause a lot of problems for companies and individuals so it's important to have security for drones.

## II. WHAT IS A DRONE?

A drone, also known as an unmanned ariel vehicle, is a small aircraft that is remotely piloted [1]. Although drones can be expensive, such as the drones used by the United States Armed Forces, many cheaper drones are available for commercial and private uses.

### A. History of Drones

The first drone created dates to 1918, during World War I. Although this drone was never actually used, its intended purpose was to deliver bomb strikes on enemy targets remotely, much as they are used to this day. Nearly fifty years later in the 1980s, drones were finally being integrated into modern combat. Today, drones are used on the battlefield for both reconnaissance and to eliminate enemy forces. Drones are also used commercially to deliver items and to take unique photos and videos [2].

### B. Drone Components

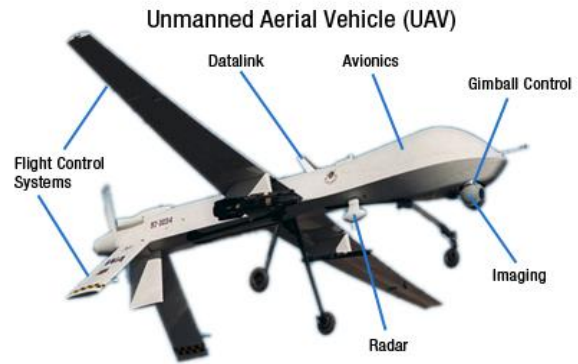

Fig. 1. A Drone [3]

Drones are composed of several elements designed to aid them in their tasks. Flight and speed controllers are used to help the drone navigate. Most drones are also equipped with cameras and data storage for surveillance purposes. Finally, drones can be either fixed-winged or use one or more rotors to achieve liftoff [2].

*A. Drone Uses*

From military drones to commercial drones, there are many ways that drones are used. First, drones are used to aid agriculture and land use such as fighting forest fires. Second, drones are used to search for and rescue people in need. Drones can also be used to monitor weather and traffic patterns to help enhance safety and response time during dangerous incidents. Another recently popular use of drones is the delivery of items from such companies as Amazon and UPS. Finally, drones can be used in military operations to keep soldiers out of harm's way [2].

### III. COMMERCIAL AND MILITARY USES

There are multiple ways that people can use drones in the modern-day world, to monitor, deliver, or in combat. Here are a couple of examples of companies that use drones in their everyday commercial infrastructure.

*A. News Industry*

Drones are being used not just for entertainment in the modern day but for legitimate business purposes. They made a big break at CNN when they were able to capture aerial coverage of the impact of the Syrian Civil War on Aleppo. It was because of this footage that they started making laws for sanctioned aerial space. Similar laws have been passed in recent years, aimed at keeping people safe as drones are integrated into everyday life [3].

*B. EasyJet*

Drones are being used in the field of aviation to make it easier for engineers to perform routine checks. The field of aviation is already at an optimal standard, as it's considered the safest form of transportation. You're more likely to be in a car accident or get struck by lightning than your chances of getting into a horrific plane crash. So, to make an already safe field safer they use drones with algorithms that can scan a plane and aren't prone to human error. Along with this, they complete checks on airbuses in hours that normally would take a couple of days or more. This allows engineers to focus on other projects which in turn saves resources [3][4].

*C. Amazon*

Amazon has been working on drone delivery for decades. They've made a prototype that's able to deliver packages that are less than five pounds. They are still working to make these drones more durable because they can't travel in certain types of weather. As well there are certain safety features that must be improved before they're able to put the actual prototypes in motion. The plan is to have these drones deliver packages by 2024 [3].

*D. Agriculture*

Drones have been in agriculture since about 1939 when they were used by the US Army to disperse seeds in Dayton Ohio.

  *a) Mapping drones*

o  Mapping Drones survey the crop field to help monitor crop health.

  *b) Seed Planters*

o  These drones' purpose is to fly over crop fields and disperse seeds over the land.

*E. Military Drones*

Unmanned Aerial Vehicles (UAV) are drones that are used by the military. The military has many different uses for drones. They are one of the biggest innovations of the century used in the military. They are essential to the military's reconnaissance as well as surveillance.

  *1) Types of Drones*

  *a) Reconnaissance Drones*

These drones are used for enemy surveillance as well as for tracking enemy movements, weapon stockpiles, and strategic targets [5].

  *b) Attack drones*

Attack drones carry weapons like missiles and bombs to carry out stealth attacks [5].

  *c) Combat Support Drones*

These drones carry weapons or survey the field to alert troops of enemy whereabouts [5].

*d) Combat Drones*

Combat drones may be equipped with machine guns, cannons, or other weapons and their purpose is to engage in air combat with enemy aircraft [5].

*e) Research and Development Drones*

These drones are used for testing equipment used in war [5].

*f) Logistics Drones*

These drones support transport operations as well as provide troops with equipment and, in special cases, can carry troops into battle [5].

*g) ISTAR (Intelligence, Surveillance, Target, Acquisition, and Reconnaissance) Drones*

In addition to the capabilities in the name, this drone can be used for electronic warfare using EMP devices or used as a decoy to throw off enemy sensors [5].

*2) Classes of Military Drones*

Class I: These are typically the slowest types of drones and are good for reconnaissance and surveillance [5]. They weigh less than 150 Kilograms.

Class II: These drones are used for air support [5]. They weigh between 150 and 600 Kilograms.

Class III: Large and fast-moving, these drones can conduct strikes from high altitudes. They serve in the highest-grade military operations [5]. They weigh over 600 Kilograms.

IV. WHY SECURITY IS IMPORTANT FOR DRONES

When you think about drones, they are nothing more than flying computers. With that in mind, drones are susceptible to being hacked. Some of the main methods a hacker can use to gain control of a drone are GPS spoofing, planting malware on the control unit of the drone, and planting malware on the drone itself.

*A. GPS Spoofing*

A hacker can use a radio transmitter to send a fake GPS signal to a receiver antenna on a drone to make the drone think that the fake GPS signal is a legitimate signal, thus making the drone return to a location of the hacker's choice [6].

*B. Malware on the Control Unit of the Drone*

Malware on the control unit of the drone can cause the user to lose control of the drone. The hacker can then take control of the infected drone and give it their own commands [8].

*C. Malware on the Drone*

Like malware on the control unit of the drone, drones themselves can also be infected with malware. This can cause a loss of control of the drone and possibly even destroy the drone so that it can no longer be used [8].

*D. Consequences of a Security Breach*

With drones being vulnerable to being hacked, it's important to have security on drones. Drones being hacked can cause serious problems for companies and individuals. In the event of a natural disaster, drones are used to conduct search and rescue because of their ability to see a wide range of an area. If a drone were to get hacked during a crucial situation like this one, it could cause lives to be lost. Drones are also used by the military, especially large combat drones. These drones are used for battlefield intelligence, reconnaissance, and surveillance. If one of these drones gets hacked during a combat mission, it would be disastrous for the military because of the amount of information the hacker could obtain. To combat these problems, drones need to have security.

V. PRIVACY AND DRONE USE

The number one issue related to a breach in drone security is the private data collected by drones could become exposed to the public. Although there are regulations about what footage and photos companies are allowed to film, there is still a chance that your private footage could be at stake, whether you consent to be filmed or not.

*A. Government Drone Use*

Governments around the world use drones to help monitor security threats and identify criminals. This includes the United States. The FBI, as well as several local law enforcement agencies, use drones to catch criminals. Although this is a good thing, many innocent people are also filmed, without their consent. It is therefore necessary to protect such private data from falling into the wrong hands [2].

*B. Commercial Drone Use*

Businesses are another way in which drone footage can be collected without permission. This could happen at an amusement park or at your home when Amazon delivers a package. Not only could this private data be stolen by hackers but the companies themselves also have access [2].

*C. Personal Drone Use*

Last but not least, private citizens can purchase and utilize drones for a variety of reasons from photography to racing. Even under the best intentions, private data can still be collected accidentally and later compromised should that drone be hacked.

*D. Compromised Data*

Why does it matter if a drone is hacked after collecting your footage? If your private data falls into the wrong hands, which is very likely in the world of technology, it can have catastrophic impacts on your life, whether it be the government or just a regular person with a drone. It is important, therefore, to maintain regulations restricting what governmental, commercial, and even private citizens can film using drones.

## VI. DRONE SECURITY

*A. Software Updates*

There are sometimes security issues and bugs that can cause vulnerability points in the drone's system. These bugs can be solved by software updates [7].

*B. Install Security Software*

Use a VPN and make sure you implement encryption in the protocols you give the drone just in case your network isn't secure, and a protocol is intercepted [7].

*C. Implement a Return Home Feature*

If all else fails and you lose connection with your drone, make a protocol for it to return to a secure location. This is an essential practice. Look at it like a PC that is shut down by a brute-force attack. You want to make sure it's shut down with the ports closed. The equivalence of shutting down the ports, in this case, would be making sure the drone returns to a secure location or self-destructs so that threats don't have access to private data [7].

*D. Make Sure to Backup Data*

Any data that is taken from the drone should be backed up to another location away from the drone, that way if an incident occurs no data is lost [7].

## VII. CONCLUSION

Drones are unmanned vehicles used to carry out a variety of operations, both commercially and privately. Many businesses, like Amazon, use drones in their everyday business operations. The United States Military also uses drones to keep soldiers out of harm's way. It is important to secure drones for two reasons. First, they can be hacked and used by the attacker. Second, data can be extracted from the drone's data storage, jeopardizing company, and personal information. Privacy will always be one of the top concerns with drones, both in the footage they capture and how that footage is kept safe from unauthorized individuals. Keeping anti-malware software up to date on a drone's system as well as having a contingency location for a drone to return to in case of an attack can help mitigate such attacks. When used responsibly with strong security practices, drones can be an amazing gift to society.